\documentclass[letterpaper, 10 pt, conference]{ieeeconf}  % Comment this line out if you need a4paper

\IEEEoverridecommandlockouts                              % This command is only needed if 
\usepackage{graphics} % for pdf, bitmapped graphics files
\usepackage{graphicx}
\usepackage{epsfig} % for postscript graphics files
\usepackage{mathptmx} % assumes new font selection scheme installed
\usepackage{times} % assumes new font selection scheme installed
\usepackage{amsmath} % assumes amsmath package installed
\usepackage{amssymb}  % assumes amsmath package installed
\usepackage{xcolor}

\title{\LARGE \bf
LUMINA-Net: Low-light Upgrade through Multi-stage Illumination and Noise Adaptation Network for Image Enhancement
}
\author{Siddiqua Namrah$^{1}$, Sun-Eung Kim$^{1}$, and Seong-Whan Lee$^{1}$% <-this % stops a space
\thanks{*This research was supported by the Institute of Information \& Communications Technology Planning \& Evaluation (IITP) grant, funded by the Korea government (MSIT) (No. RS-2019-II190079 (Artificial Intelligence Graduate School Program (Korea University)), and No. IITP-2025- RS-2024-00436857 (Information Technology Research Center (ITRC)).}% <-this % stops a space
\thanks{$^{1}$S. Namrah, S.-E. Kim, and S.-W. Lee are with the Department of Artificial Intelligence, Korea University, Anam-dong, Seongbuk-ku, Seoul 02841, Korea.
{\texttt{\small \{namrah96, se\_kim, sw.lee\}@korea.ac.kr}}}}

\begin{document}
\maketitle
\thispagestyle{empty}
\pagestyle{empty}

\begin{abstract}
Low-light image enhancement (LLIE) is a crucial task in computer vision aimed at enhancing the visual fidelity of images captured under low-illumination conditions. Conventional methods frequently struggle with noise, overexposure, and color distortion, leading to significant image quality degradation. To address these challenges, we propose LUMINA-Net, an unsupervised deep learning framework that learns adaptive priors from low-light image pairs by integrating multi-stage illumination and reflectance modules. To assist the Retinex decomposition, inappropriate features in the raw image can be removed using a simple self-supervised mechanism. First, the illumination module intelligently adjusts brightness and contrast while preserving intricate textural details. Second, the reflectance module incorporates a noise reduction mechanism that leverages spatial attention and channel-wise feature refinement to mitigate noise contamination. Through extensive experiments on LOL and SICE datasets, evaluated using PSNR, SSIM, and LPIPS metrics, LUMINA-Net surpasses state-of-the-art methods, demonstrating its efficacy in low-light image enhancement.
\end{abstract}

\section{Introduction}
Low-Light Image Enhancement (LLIE) has emerged as a vital component in various image-based applications, including surveillance \cite{qu2024double,liu2022attention}, autonomous vehicles \cite{li2024light,liu2024lane}, medical imaging \cite{ma2021structure,yang2007reconstruction}, and consumer electronics \cite{fu2022efficient,zhou2024real,lee1999integrated}, where high-fidelity images are paramount. Traditional approaches to LLIE have primarily relied on two well-established methods, histogram-based and retinex-based techniques. Histogram-based LLIE techniques analyze and modify pixel intensity distributions to adjust contrast and brightness, with Histogram Equalization (HE) being a widely used method \cite{banik2018contrast,park2022histogram,lee2001automatic}. Retinex-based LLIE techniques separate images into reflectance and illumination components, adjusting the latter to enhance contrast and visibility while preserving natural colors and textures \cite{lee1996multiresolution,hai2023r2rnet,lee2020uncertainty}.

\begin{figure}[thpb]
    \includegraphics[width=7.cm]{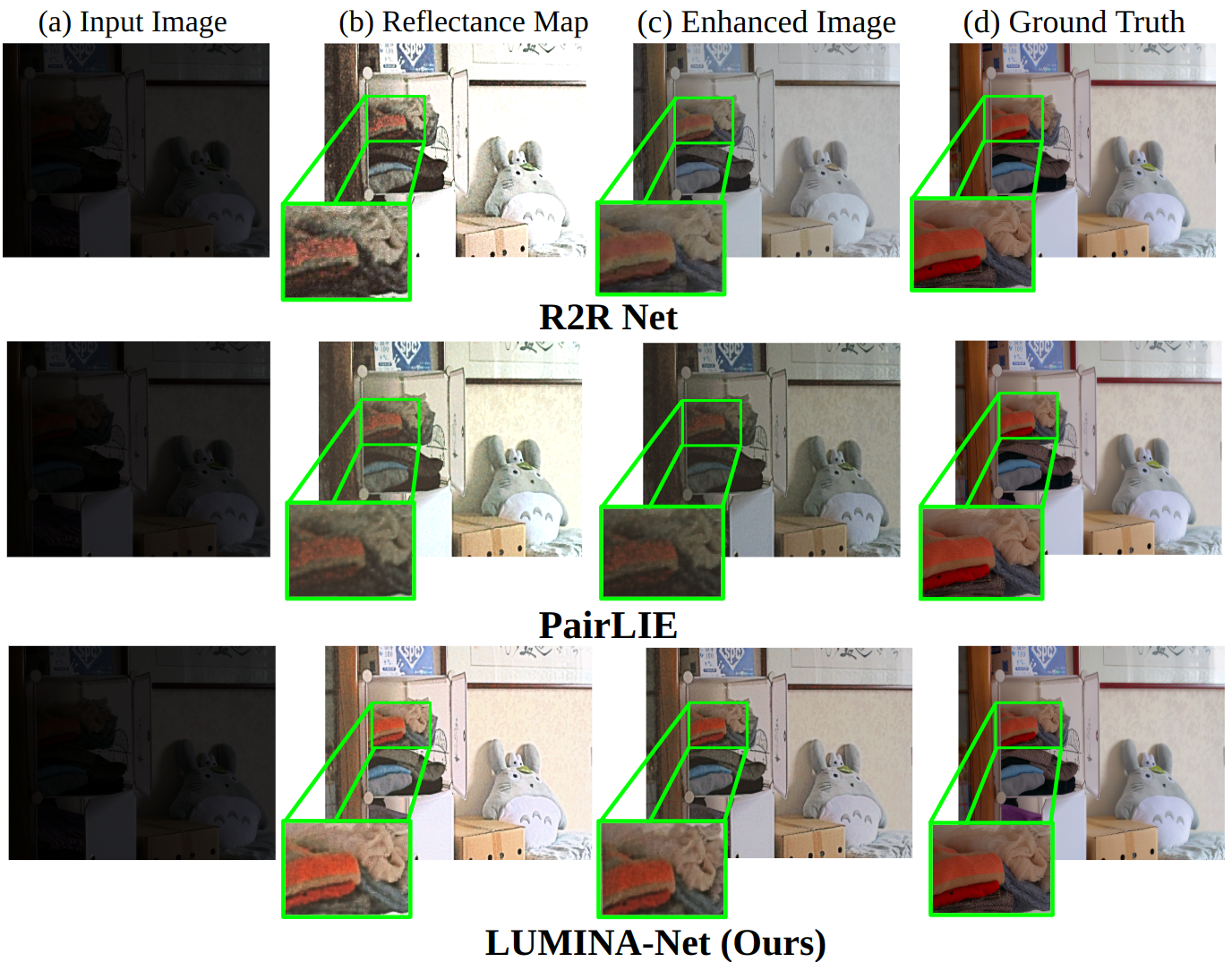}
    \centering
    \caption{A comparative analysis of reflectance images utilized in existing methods and their corresponding final results reveals significant limitations. These limitations are evident across four scenarios: (a) input images captured under low-light conditions, (b) reflectance maps that frequently amplify overexposure artifacts, (c) enhanced images prone to color distortions and texture loss, and (d) ground truth images showcasing the desired balance of color fidelity and structural detail. }
    \label{figure1}
\end{figure}

However, conventional image capture and processing methods frequently fall short in low-light conditions, underscoring the urgent need for groundbreaking solutions that can effectively mitigate the challenges of diminished illumination. Existing Retinex-based methods rely on single-image inputs, limiting their ability to address varying exposure levels effectively.

The comparative analysis in Fig. \ref{figure1} highlights the inherent limitations of these approaches.  Specifically, it reveals that overexposure remains a critical challenge, significantly degrading the quality of the reconstructed images. As depicted in Fig. \ref{figure1}, reflectance images processed by existing methods often suffer from pronounced artifacts caused by overexposure, leading to imbalanced color distribution and loss of texture detail. This inadequacy in handling under-exposed and over-exposed regions results in outputs marred by color distortion and diminished visual fidelity.

These challenges highlight the need for advanced, adaptive methods that handle diverse exposure conditions, reduce noise, and mitigate overexposure—ultimately enhancing image quality while preserving texture and color accuracy in difficult lighting.
%%FIGURE2
\begin{figure*}[h]
    \centering
    \includegraphics[width=0.82\textwidth]{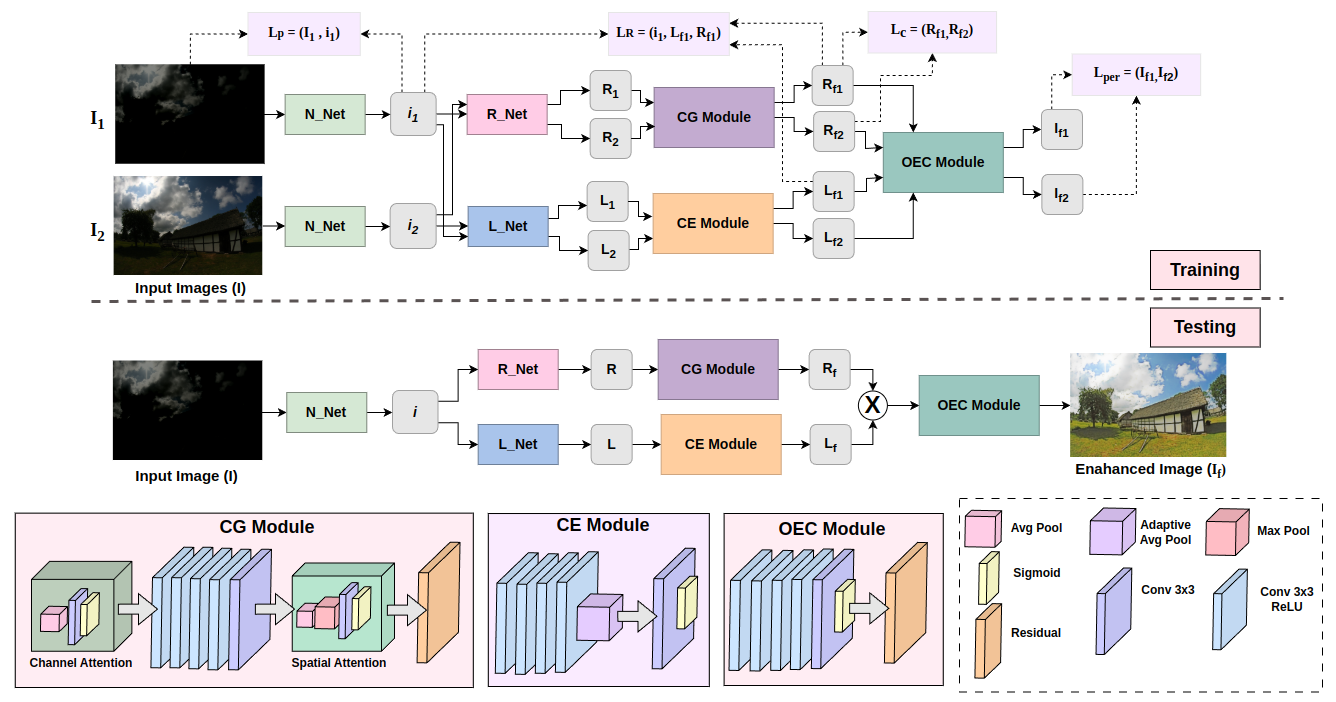}
    \caption{The proposed LUMINA-Net consists of three modules: the N-Net, R-Net, and L-Net for initial image decomposition and processing, followed by the CG Module for channel-guided reflectance enhancement, the CE Module for illumination cross-enhancement, and the OEC Module for over-exposure correction. In Training Phase Input images \( I_1 \) and \( I_2 \) are first passed through N-Net and the refined images are \(i_1, i_2 \) which are then decomposed into reflectance \(R_1, R_2 \) and illumination \( L_1, L_2 \). In Testing Phase final enhanced image \( I_f \) is produced after combining illumination and reflectance components, ensuring improved lighting, reduced noise, and preserved structural details. Multiple loss functions, including perceptual loss \( L_{per} \), Consistency Loss \(L_c \), Reflectance Loss \(L_R \), and Projection loss \( L_{p} \), are employed to optimize the enhancement quality.}
    \label{figure2}
\end{figure*}

To address the challenges of low-light image enhancement, we propose LUMINA-Net, a deep learning framework that utilizes paired low-light images with varying exposures. Inspired by Retinex theory, LUMINA-Net decomposes images into illumination and reflectance components. A key feature is the CG Module, which uses spatial and channel attention to refine feature extraction. The CE and OEC Modules ensure natural color reproduction and dynamic adjustment of overexposed regions, reducing artifacts. Additionally, self-attention-based skip connections preserve global coherence and fine details.

Despite progress in deep learning techniques like Generative Diffusion Prior (GDP) \cite{fei2023generative}, unsupervised denoising \cite{lin2023unsupervised}, and noise-mitigation networks \cite{cao2024zero}, overexposure and noise remain persistent challenges, often degrading image quality. Traditional methods like Retinex-based or histogram-based approaches do not fully address these issues. LUMINA-Net aims to overcome these limitations, improving both visual quality and system performance under challenging lighting conditions.

\section{Method}
We define the problem of LLIE using paired low-light images and emphasize the need for an effective enhancement approach. LUMINA-Net employs a structured pipeline that refines illumination, restores color accuracy, and balances exposure. The network enhances visibility while preserving fine details and reducing unwanted distortions, ensuring natural-looking results. Carefully designed loss functions guide optimization, improving stability and overall image quality. Finally, we introduce the training dataset, which includes diverse lighting conditions to enhance model generalization. The overall network structure is illustrated in Fig. \ref{figure2}.

%PAIRED IMAGES
\subsection{Preliminary}
According to the Retinex theory, low-light image $I$ can be decomposed into illumination L and reflectance R as, 
\begin{equation}
    I = L \circ R ,
    \label{Eq1}
\end{equation}
where $\circ$ denotes element-wise multiplication. Illumination $(L)$ represents the light intensity in the scene, expected to be smooth and texture-less, while reflectance $(R)$ captures the inherent properties of objects, such as textures and details. Conventional illumination and reflectance estimation techniques rely on predefined, hand-crafted priors that often do not accommodate the complexity and variability of real-world scenes and lighting conditions \cite{sun2024di, sethu2023comprehensive}. To address this limitation, we utilize paired low-light images, $I_1$ and $I_2$ with same reflectance $R'$ but different illuminations $L_1'$ and $L_2'$.
\begin{equation}
    I_1 = L_1' \circ R', \quad I_2 = L_2' \circ R' .
    \label{Eq2}
\end{equation}

With paired low-light instances, we propose LUMINA-Net that injects additional constraints and valuable information, bolstering the robustness of illumination-reflectance decomposition.
\subsection{Proposed Method}
Our proposed LUMINA-Net image enhancement technique seamlessly integrates three synergistic components, the CG Module for adaptive illumination-reflectance decomposition, the CE Module for vibrant color restoration and refinement, and the OEC Module for balanced brightness and detail preservation. This combination enables effective low-light image enhancement by improving brightness, reducing noise, and maintaining structural fidelity across diverse scenes.
%CGNET
\subsubsection{\textbf{Channel-Spatial Guidance (CG) Module}} 
The CG Module plays a crucial role in refining reflectance maps generated from the illumination-reflectance decomposition. It employs both Channel Attention and Spatial Attention mechanisms to selectively highlight important features and suppress noise, ensuring better quality reflectance maps.

\textbf{Channel Attention Mechanism} works by first applying global average pooling (GAP) to the reflectance maps \(R_1\) and \(R_2\) to generate channel descriptors. These descriptors are refined via a convolutional layer and sigmoid activation to produce channel-wise weights, which prioritize essential features in the reflectance maps \cite{hu2018squeeze}.

\textbf{Spatial Attention Mechanism} focuses on critical spatial regions within the reflectance maps, such as edges and textures. This is achieved through global average pooling and max pooling along the channel dimension, generating spatial descriptors. These descriptors are then processed through a convolutional layer and sigmoid activation to produce a spatial weight map that highlights important regions for further refinement \cite{woo2018cbam}.

By combining both attention mechanisms, the CG Module generates a more refined reflectance map from \(R_{f1}\) and \(R_{f2}\), improving the accuracy and detail preservation of the enhanced image while reducing noise. This results in a cleaner and more precise representation of the reflectance, outperforming traditional Retinex-based approaches.

\subsubsection{\textbf{Color Enhancement (CE) Module}} 
The CE Module refines the illumination maps \(L_1\) and \(L_2\) to improve brightness and contrast, especially in low-light regions, while maintaining natural lighting transitions.

It uses convolutional layers to enhance the illumination maps and Adaptive Average Pooling (AAP) to generate global channel descriptors. These descriptors are processed through a fully connected layer and sigmoid activation to produce channel-wise attention weights, emphasizing important features and suppressing irrelevant ones \cite{zhang2023adaptive}.

The refined illumination maps \(L_{f1}\) and \(L_{f2}\) are derived by aggregating spatial information through Adaptive Average Pooling (AAP), enhancing the illumination estimates. This process, driven by attention mechanisms, ensures more accurate and visually pleasing results in both low-light and normal-light conditions.

%OCEM
\subsubsection{\textbf{Over-Exposure Correction (OEC) Module}}  
The OEC Module is designed to address the issue of overexposure by refining the combined illumination $(L_f)$ and reflectance $(R_f)$ maps. This process restores details in overexposed regions, ensuring a balanced exposure through an intermediate image representation, generated by element-wise multiplying the refined illumination and reflectance map. This intermediate representation is then passed through the OEC Module, which is specifically designed to handle areas of the image that are excessively bright or saturated.  
To compute the final enhanced image, we apply the illumination correction factor $\lambda$ as follows:  
\begin{equation}  
I_{f} = L^\lambda \circ R,
\label{Eq3}
\end{equation}  
where $\lambda$ is the illumination correction factor, and $I_{f}$ denotes the enhanced image. This ensures a refined and well-balanced output by dynamically adjusting the illumination to prevent excessive brightness while maintaining structural details.  

%LOSS
\subsection{Loss Functions}
The LUMINA-Net architecture employs a multi-faceted loss function strategy to ensure effective training and image enhancement. This integrated approach minimizes differences between predicted and ground-truth images while preserving critical image characteristics, including perceptual quality, spatial smoothness, and reflectance consistency, thereby achieving a harmonious balance between visual fidelity and structural integrity.
%Projection Loss
\subsubsection{\textbf{Projection Loss($L_p$)}}  The projection step ensures that the input image is more suitable for decomposition under the Retinex model by removing noise and irrelevant features. The projection loss measures the difference between the original input image $I_1$ and the projected image $i_1$, guiding the transformation of the original image into a cleaner, noise-free representation that better aligns with the Retinex assumptions. This loss can be expressed as:
\begin{equation}
    L_p = ||I_1 - i_1||_2^2 .
    \label{Eq4}
\end{equation}

By reallocating the decomposition error to the projection stage, this process ensures more accurate and detailed reflectance and illumination maps, resulting in enhanced image quality and realistic reconstructions.
%Consistency Loss
\subsubsection{\textbf{Consistency Loss ($L_C$)}} The $L_C$ is derived from the Retinex theory and plays a pivotal role in maintaining consistency between the reflectance maps of paired low-light images. This consistency enforces accurate reflectance decomposition and implicitly addresses sensor noise without requiring additional handcrafted constraints. The loss is defined as:
\begin{equation}
    L_C = ||R_{f1} - R_{f2}||_2^2 ,
    \label{Eq5}
\end{equation}
where $R_{f1}$ and $R_{f2}$ are the reflectance components of the paired low-light images.

By minimizing the difference between the reflectance maps of the two images, $L_C$   leverages the randomness of noise across paired images, enabling effective noise removal while ensuring reflectance consistency. This enhances the robustness and accuracy of the decomposition process for low-light image enhancement.
%Retinex Loss
\subsubsection{\textbf{Retinex Loss ($L_R$)}}This approach decomposes low-light images into illumination and reflectance components. The goal is to estimate the reflectance map, capturing intrinsic scene properties, and the illumination map, representing lighting conditions. Constraints are applied to ensure the decomposition is consistent and physically meaningful, resulting in high-quality enhancements. The loss function is formulated as follows:
\begin{align*}
    L_R = ||R_{f1} \circ L_{f1} - i||_2^2 
    \quad + ||R_{f1} - i/stopgrad(L_{f1})||_2^2
\end{align*}
\begin{equation}
    \quad + ||L - L_0||_2^2 + ||\nabla L||_1 ,
\label{Eq6}
\end{equation}
where \(i\) is the input low‑light image, \(R_{f1}\) and \(L_{f1}\) are the predicted reflectance and illumination maps, \(L_0\) is the initial illumination estimate, and \(\nabla L\) represents the illumination gradient. The operator \(\mathrm{stopgrad}(\cdot)\) acts as an identity in the forward pass but blocks gradient flow during backpropagation, keeping \(L_{f1}\) fixed in the term \(R_{f1} - i / \mathrm{stopgrad}(L_{f1})\). This prevents trivial decompositions by focusing training on \(R_{f1}\). Finally, \(\|L - L_0\|_2^2\) anchors the illumination map, and \(\|\nabla L\|_1\) promotes smoothness. Minimizing this loss effectively separates the reflectance and illumination components, enhancing the low‑light image quality.

%Table1
\begin{table*}[ht]
    \centering
    \small
    \scriptsize
    \caption{\centering\scriptsize QUANTITATIVE COMPARISONS WITH STATE-OF-THE-ART METHODS ON LOL AND SICE DATASETS. “T”, “S”, and “U” REPRESENT “TRADITIONAL", “SUPERVISED", AND “UNSUPERVISED" METHODS, RESPECTIVELY. THE BEST, SECOND AND THIRD PERFORMANCES ARE MARKED IN \textcolor{red}{RED}, \textcolor{blue}{BLUE}, AND \textcolor{green}{GREEN}, RESPECTIVELY.}
    \scriptsize
    \resizebox{0.55\textwidth}{!}{%
    \begin{tabular}{l@{\hskip 10pt}|c@{\hskip 10pt}|ccc@{\hskip 10pt}|ccc}
    \hline
    \textbf{Method}        & \textbf{Type} & \multicolumn{3}{c|}{\textbf{LOL}} & \multicolumn{3}{c}{\textbf{SICE}} \\ 
    \cline{3-8}
     & & \textbf{PSNR↑} & \textbf{SSIM↑} & \textbf{LPIPS↓} & \textbf{PSNR↑} & \textbf{SSIM↑} & \textbf{LPIPS↓} \\ 
    \hline
    SDD \cite{hao2020low}& T & 13.34 & 0.637 & 0.743 & 15.35 & 0.741 & 0.232\\ 
    STAR \cite{xu2020star} & T & 12.91 & 0.518 & 0.366 & 15.17 & 0.727 & 0.246\\ 
    \hline
    MBLLEN \cite{lv2018mbllen} & S & 17.86 & 0.727 & 0.225 & 13.64 & 0.632 & 0.297\\ 
    RetinexNet \cite{wei2018deep} & S & 17.55 & 0.648 & 0.379 & 19.89 & 0.783 & 0.276\\ 
    GLADNet \cite{wang2018gladnet} & S & 19.72 & 0.680 & 0.321 & 19.98 & 0.837 & 0.203\\ 
    KinD \cite{zhang2019kindling} & S & 17.65 & {0.775} & \textcolor{blue}{0.171} & \textcolor{blue}{21.10} & 0.838 & \textcolor{green}{0.195}\\ 
    DRBN \cite{yang2020fidelity} & S & 16.29 & 0.551 & 0.260 & 15.58 & 0.522 & 0.289\\
    URetinexNet \cite{wu2022uretinex} & S & {19.84} & \textcolor{red}{0.826} & \textcolor{red}{0.128} & \textcolor{blue}{21.64} & \textcolor{green}{0.843} & \textcolor{blue}{0.192}\\ 
    \hline
    ZeroDCE \cite{guo2020zero} & U & 14.86 & 0.559 & 0.335 & 18.69 & 0.810 & 0.207\\ 
    RRDNet \cite{zhu2020zero} & U & 11.40 & 0.457 & 0.362 & 13.28 & 0.678 & 0.221\\ 
    RUAS \cite{liu2021retinex} & U & 16.40 & 0.500 & 0.270 & 16.85 & 0.734 & 0.363\\ 
    SCI \cite{ma2022toward} & U & 14.78 & 0.522 & 0.339 & 15.95 & 0.787 & 0.235\\ 
    EnlightenGAN \cite{jiang2021enlightengan} & U & 17.48 & 0.651 & 0.322 & 18.73 & 0.822 & 0.216\\ 
    PairLIE \cite{fu2023learning} & U & 19.51 & 0.736 & 0.248 & 21.32 & \textcolor{blue}{0.840} & 0.216\\ 
    NeRco \cite{yang2023implicit} & U & {19.80} & 0.730 & 0.240 & 20.73 & 0.820 & 0.230\\ 
    FourierDiff \cite{lv2024fourier} & U & \textcolor{blue}{21.94} & 0.710 & 0.362 & 18.87 & 0.768 & 0.387\\ 
    LightenDiffusion \cite{jiang2024lightendiffusion} & U & \textcolor{green}{20.45} & \textcolor{green}{0.803} & 0.209 & 19.08 & 0.776 & 0.375\\ 
    LUMINA-Net (Ours) & U &\textcolor{red}{23.92} & \textcolor{blue}{0.812} & \textcolor{green}{0.180} & \textcolor{red}{22.65} & \textcolor{red}{0.853} & \textcolor{red}{0.131}\\ 
    \hline
\end{tabular}}
\label{tab:comparison}
\end{table*}
%Figure3
\begin{figure*}[h]
\centering
\includegraphics[width=0.60\textwidth]{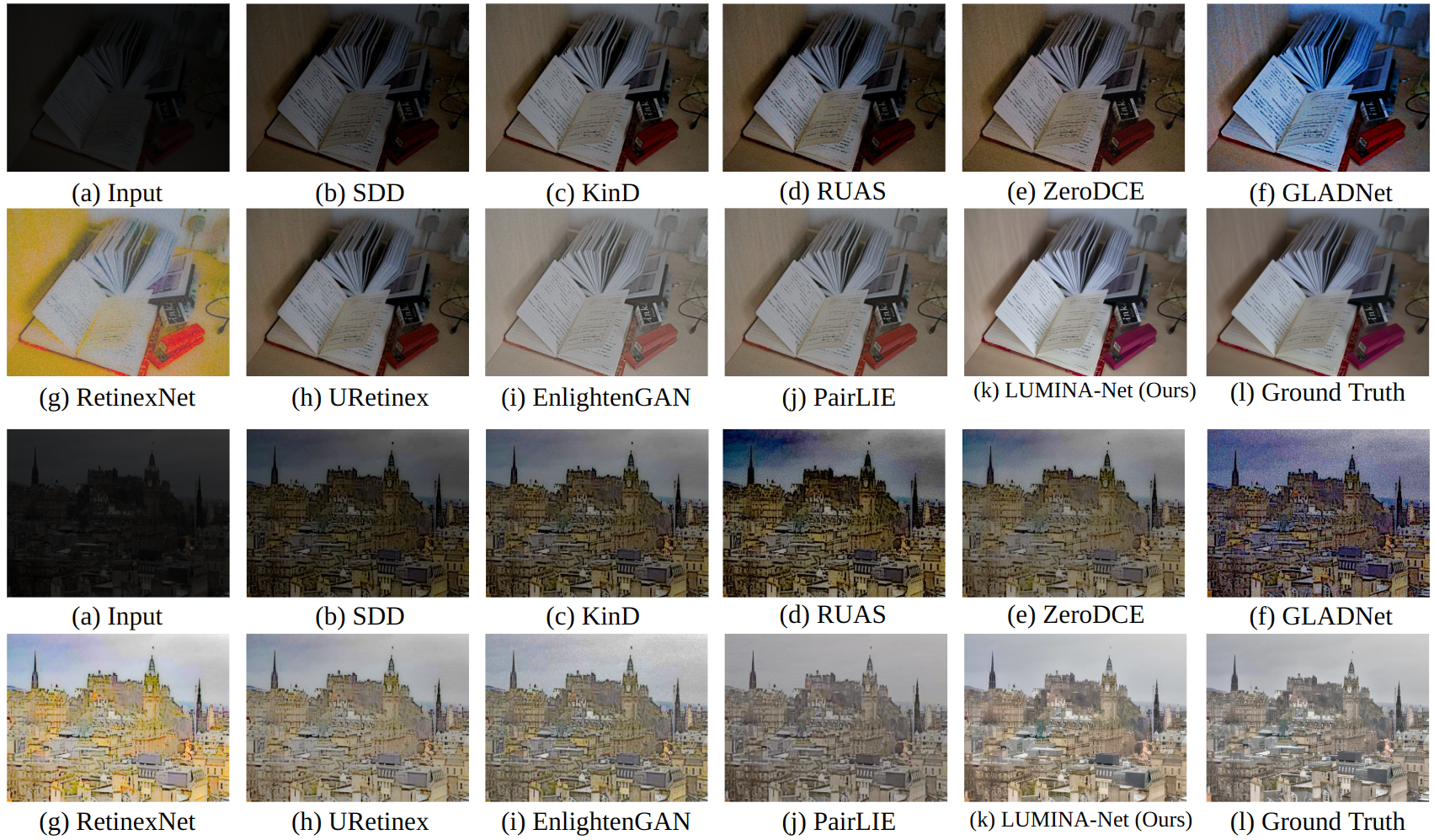}
\caption{Visual comparisons of different low-light image enhancement (LIE) methods. (a) Input image. (b-j) Results from various state-of-the-art methods. (k) Our proposed method, LUMINA-Net, demonstrating superior enhancement. (l) Ground truth image for reference. The images (b-j) show the performance of existing techniques, while (k) highlights the effectiveness of LUMINA-Net in restoring details and colors in low-light conditions.}
\label{figure3}
\end{figure*}

%PerceptualLoss
\subsubsection{\textbf{Perceptual Loss ($L_{per}$)}} The $L_{per}$ measures the similarity between the two predicted enhanced images ($I_{f1}$ and $I_{f2}$) in a high-level feature space, capturing their perceptual consistency. This encourages both enhanced outputs to retain visually coherent features by comparing their representations extracted from a pre-trained network.
\begin{equation}
    L_{per} = ||(\phi(I_{f1}), \phi(I_{f2}))||_2^2 ,
    \label{Eq7}
\end{equation}
where $I_{f1}$ and $I_{f2}$ are the predicted enhanced images from different augmented inputs, and \( \phi(\cdot) \) represents the feature extraction function of the pretrained model.

%Combined Loss
\subsection{Combined Loss} The final loss is a weighted sum of individual losses, optimizing perceptual, consistency, projection, reflectance, and edge losses to balance visual fidelity, detail, and spatial coherence. The Combined Loss is formulated as follows:
\begin{align*}
    L_{\text{All}} = w_0 \times L_p + w_1 \times L_{C} + w_2 \times L_{R} + w_3 \times L_{per},
\end{align*}
where \(L_{p}\) is projection loss, \(L_{C}\) is reflectance consistency loss, \(L_{R}\) is retinex loss, and \(L_{per}\) is perceptual loss, with \(w_0, w_1, w_2, w_3\) as weights.
%Figure4
\begin{figure*}[t]
\centering
\includegraphics[width=0.65\textwidth]{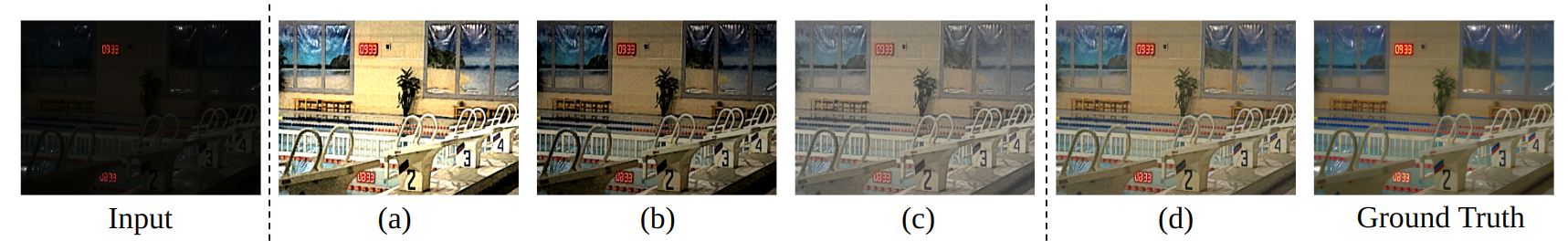}
\caption{Visual comparisons of the ablation studies. (a) Without the OEC Module. (b) Without the CG Module. (c) Without the CE Module. Our result (d) demonstrates the most visually accurate restoration, closely matching the ground truth. The ablation studies highlight the contribution of each module in improving the overall enhancement performance.}
\label{figure4}
\end{figure*}

\section{Experiments}
    \subsection{Experimental Datasets}
     LUMINA-Net is trained using low-light image pairs derived from the SICE \cite{cai2018learning} and LOL datasets \cite{wei2018deep}. For evaluation, we select an additional 50 sequences (150 images) from the SICE dataset and utilize the official evaluation set (15 images) from the LOL dataset to assess the model’s performance. Both SICE and LOL contain reference images, allowing us to employ multiple metrics for objective evaluation, including PSNR, SSIM \cite{wang2004image}, LPIPS \cite{zhang2018unreasonable}. A higher PSNR or SSIM score indicates that the enhanced result is closer to the reference image in terms of fidelity and structural similarity. Conversely, lower LPIPS values signify improved enhancement quality and more accurate color reproduction. %Additionally, we use the MEF dataset \cite{ma2015perceptual} for qualitative visual comparisons, further validating the effectiveness of LUMINA-Net against other methods.

     \subsection{Implementation Details}
     We implement LUMINA-Net using PyTorch. During training, images are randomly cropped to \(256 \times 256\), and a batch size of 1 is used for efficient memory usage. The ADAM optimizer \cite{adam2014method} is employed with an initial learning rate of \(1 \times 10^{-4}\), along with a cosine annealing scheduler. Training runs for 400 epochs for convergence. The default correction factor is \(\lambda = 0.2\), as in \cite{fu2023learning}, and adjusted to \(\lambda = 0.10\) for the LOL dataset. The loss weights \(w_0 = 5\), \(w_1 = 1\), \(w_2 = 1\), and \(w_3 = 0.1\) are set based on empirical evaluation. Models are run on NVIDIA TITAN XP GPUs.
     
     \subsection{Comparison with state-of-the-arts methods}
     LUMINA-Net is compared with 16 state-of-the-art low-light image enhancement (LIE) methods, which can be grouped into three categories: traditional methods (SDD \cite{hao2020low}, STAR \cite{xu2020star}), supervised approaches (MBLLEN \cite{lv2018mbllen}, RetinexNet \cite{wei2018deep}, GALDNet \cite{wang2018gladnet}, KinD \cite{zhang2019kindling}, DRBN \cite{yang2020fidelity}, URetinexNet \cite{wu2022uretinex}), and unsupervised methods (Zero-DCE \cite{guo2020zero}, RRDNet \cite{zhu2020zero}, RUAS \cite{liu2021retinex}, SCI \cite{ma2022toward}, EnlightenGAN \cite{jiang2021enlightengan}, PairLIE \cite{fu2023learning}, FourierDiff \cite{lv2024fourier}, and NeRco \cite{yang2023implicit}). These comparisons are made based on the performance of each method, using their official codes with the recommended parameters to ensure fairness and consistency in the evaluation. The results obtained from these methods are used as a benchmark for assessing LUMINA-Net’s performance.

     \subsection{Quantitative Comparisons}
     Table \ref{tab:comparison} presents the quantitative performance results on the LOL and SICE datasets. Traditional and unsupervised methods show relatively poor performance, as they face challenges in learning enhancement models without reference images. Traditional methods, relying on fixed algorithms, struggle to adapt to varied lighting conditions, while unsupervised methods, lacking paired references during training, depend on indirect cues, limiting their effectiveness in diverse scenarios.
     
      \subsection{Visual comparisons}
     Fig. \ref{figure3} compares various low-light image enhancement methods on the LOL-real and SICE datasets. LUMINA-Net outperforms other methods by producing natural, balanced results with accurate brightness, color, and contrast. In contrast, RetinexNet suffers from overexposure, washing out details and disrupting color reproduction, leading to lower performance in metrics like SSIM and LPIPS.

    LPIPS (Learned Perceptual Image Patch Similarity) is a key metric that captures perceptual similarity using deep neural network features. Unlike PSNR and SSIM, LPIPS better aligns with human perception, exposing overexposure issues in methods like RetinexNet. LUMINA-Net achieves lower LPIPS scores, reflecting closer visual similarity to the ground truth.

    In contrast, methods like ZeroDCE and RUAS struggle with dark regions and noise, while KinD and URetinexNet lack robustness under varied lighting. LUMINA-Net outperforms across all metrics by preserving details, enhancing structure, and delivering high perceptual quality.

    \subsection{Ablation Studies}
    To validate the effectiveness of LUMINA-Net's components, including its modules and overall design, we conducted ablation experiments on the LOL dataset, with results presented in Table \ref{tab:ablation_lol} and Fig. \ref{figure4}.

%Table2
%\setlength{\tabcolsep}{12pt}
\begin{table}[ht]
\small
\centering
\caption{QUANTITATIVE RESULTS OF ABLATION STUDIES ON LOL DATASET. THE BEST RESULTS ARE MARKED IN \textbf{BOLD}.}
\resizebox{0.23\textwidth}{!}{  % Resize to 80% of text width
\begin{tabular}{lcc}
\hline
\textbf{Method} & \textbf{PSNR$\uparrow$} & \textbf{SSIM$\uparrow$} \\
\hline
w/o OEC    & 19.70  & 0.646 \\
w/o CG     & 21.50  & 0.622 \\
w/o CE     & 22.69  & 0.782 \\
Ours       & \textbf{23.92}  & \textbf{0.812} \\
\hline
\end{tabular}
}
\label{tab:ablation_lol}
\end{table}

    In Fig. \ref{figure4} (a), removing the OEC Module caused overexposure and uneven brightness, significantly lowering PSNR and SSIM scores, emphasizing its importance in brightness management, (b) shows that excluding the CG Module resulted in a loss of reflectance detail and structural consistency, underlining its role in reflectance-illumination separation for sharper details, (c), removing the CE Module led to color distortions and reduced vibrancy, highlighting its importance in color accuracy. Finally, (d), our enhanced result, demonstrates the full LUMINA-Net configuration, showing preserved fine details, accurate colors, and balanced exposure. The results confirm that OEC, CG, and CE Modules are crucial for achieving high-quality low-light image enhancement, with reflectance enhancement being essential for maintaining sharpness and detail. 

\section{Conclusion}
In this paper, we introduce LUMINA-Net to address challenges in low-light image enhancement, such as preserving details in dark regions and accurately recovering colors. It integrates three modules: CG for reflectance enhancement, CE for color restoration, and OEC for overexposure correction. LUMINA-Net improves image quality and naturalness, outperforming existing methods in detail preservation, color accuracy, and artifact reduction.

% The preferred spelling of the word ``acknowledgment'' in America is without 
% an ``e'' after the ``g''. Avoid the stilted expression ``one of us (R. B. 
% G.) thanks $\ldots$''. Instead, try ``R. B. G. thanks$\ldots$''. Put sponsor 
% acknowledgments in the unnumbered footnote on the first page.
\bibliographystyle{IEEEtran}
\bibliography{root}  % Your .bib file containing references

\end{document}